# A Reference-less Slope Detection Technique in 65nm for Robust Sensing of 1T1R Arrays

Seyedhamidreza Motaman[1], Swaroop Ghosh[2], Jae-Won Jang, Anirudh Iyengar, Rekha Govindaraj[#] and Zakir Khondker*
Computer Science and Engineering, Pennsylvania State University, [#]University of South Florida *Intel Labs
[1]sxm844@psu.edu, [2]szg212@psu.edu

***Abstract-- Spin-Torque-Transfer RAM (STTRAM) is a promising technology however process variation poses serious challenge to sensing. To eliminate bit-to-bit process variation we propose a reference-less, destructive slope detection technique which exploits the MTJ switching from high to low state to detect memory state. A proof-of-concept fabricated test-chip using 96kb mimicked STTRAM bits in 65nm technology shows that slope sensing reduces failure rate by 120X in 2.5K-5K array@TMR=100% and 162X in 2.5K-5K@TMR=80% array compared to conventional voltage sensing.***

## I. INTRODUCTION

Spin-Torque-Transfer RAM (STTRAM) is a promising memory technology for embedded cache due to high-density, low standby power and high speed. One of the crucial challenges in STTRAM is poor Sense Margin (SM). The sense margin of STTRAM depends on TMR (Tunnel Magneto Resistance) which is defined as 100*($R_H$-$R_L$/$R_L$) where $R_L$ and $R_H$ are low and high resistance of MTJ respectively. Due to poor TMR, the voltage/current differential between $R_H$ and $R_L$ decreases which degrades the SM.

STTRAM sensing can be broadly categorized into destructive and non-destructive sensing. A non-destructive voltage sensing and a sizing methodology to improve the SM of MRAM arrays has been proposed in [1]. Source degeneration is proposed [2] to reduce large sense margin variation. Negative resistance read and write technique is presented [3] to eliminate read disturb and reduce the write power. However, these sensing schemes suffer from reference resistance variation. In [4], a non-destructive self-reference sensing scheme has been proposed by leveraging the dependency of high and low resistance state of the MTJ on the cell current amplitude. Even though it reduces the read latency and power by eliminating two write steps, the sense margin is much smaller than destructive self-reference scheme and conventional non-destructive voltage sensing. Under destructive sensing, a self-reference sensing has been proposed in [5] to eliminate bit-to-bit process variation in MTJ resistance. However, it suffers from unoptimized selection of data and reference current. A short pulse reading scheme to reduce sensing time with 0% read disturbance is proposed in [21]. However, it does not focus on reducing the sensing failures.

*Proposed Idea:* If the MTJ resistance is low, it will only switch with a negative current. The resistance will remain low for positive current. We also note that slope of MTJ V-I curve changes from positive to negative during switching of resistance. Therefore, we propose to sense the change in slope of voltage to detect the MTJ state. A ramp current is injected into bitcell which results in a ramp voltage. If the MTJ resistance state is initially high, the slope of voltage will change from positive to negative as the MTJ resistance switches from high to low resistance state. The voltage slope will remain positive if the resistance of MTJ is low initially. Therefore, sensing problem can be simplified to slope detection. If a negative slope is detected then the data is sensed as '1', else the data is sensed as '0'. We used high speed sample and hold circuit to detect the slope of voltage across bitcell (Fig. 1(a). Slope sensing is destructive in nature and the value of bitcells which are in high resistance state (storing '1') must be restored after the read operation.

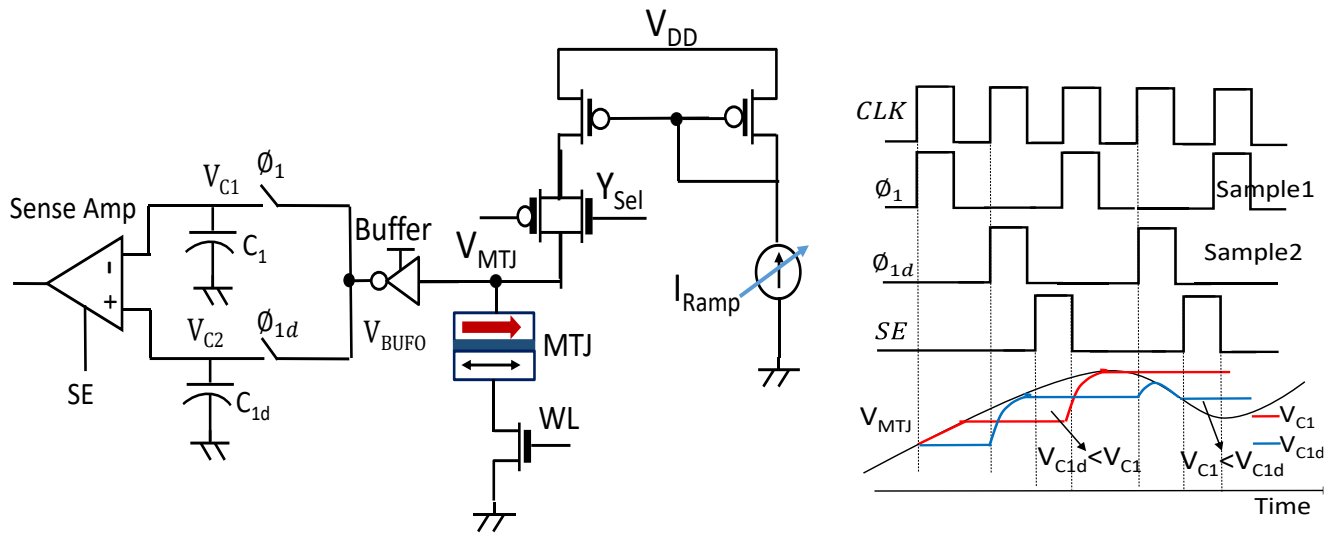

(a) (b)

Fig. 1(a) Slope detection sense circuit; and, (b) simplified timing diagram.

In this paper, we design a proof-of-concept test-chip using 96kb mimicked STTRAM (using passive resistors) bits in 65nm technology to validate the proposed slope sensing circuit. The resistor values are matched with the experimentally calibrated simulated models to capture the process variations in real MTJ. A single mimicked STTRAM bitcell contains both low and high resistors in parallel and the switching circuit is designed to match the switching latency of real MTJ.

Compared to [13], we make following additional contributions in this paper:

- We have implemented the conventional non-destructive voltage sensing with source degeneration which contains the array of 2.5K-5K and 5K-10K resistances to study the effect of resistance value on sensing failures.
- We incorporated test features to modulate the TMR and clamp voltage to investigate their effect on sensing failure.

- We have performed process variation analysis to comprehensively characterize the behavior of MTJ under process variation in terms of switching time and resistance variation and mimicked the MTJ using poly resistance.
- We have designed a test-chip and performed thorough characterization.
- We have investigated the MTJ endurance for slope sensing method and compared it against that of conventional voltage sensing. Our analyses indicate that slope sensing can achieve an endurance of ~$2.1*10^{16}$ which is sufficient to guarantee 5-7 years lifetime of computing systems.

The paper is organized as follows. In Sections II and III, we present the details of slope sensing technique and test prototype design. Section IV describes the experimental results. Process variation analysis and comparative analysis is also performed. The MTJ endurance and write-back operation are discussed in Section V. Conclusions are drawn in Section VI.

## II. STTRAM Sensing Schemes

In this section, first we describe the conventional sensing technique and its challenges. Next, we describe the proposed slope sensing technique. In addition, we propose double sampling technique to improve the robustness of slope sensing.

### A. Conventional Sensing [1-2]

The purpose of the sense circuit is to identify the resistance of the data MTJ. In order to make the comparison, data MTJ resistance is compared against reference MTJ resistance. Fig. 2(a) shows the typical voltage sensing circuit where a reference current is injected in both data and reference legs and the resulting voltage is compared by a voltage sense amplifier. Poor sense margin can result in wrong interpretation of the MTJ state.

Two critical transistors in STTRAM sense circuit are the PMOS load (PL) and NMOS clamp (NC) (Fig. 2(b)). The clamp voltage and clamp transistor size set the current in the leg (Fig. 2(b). The load transistor sets the output voltage (where the NMOS and PMOS drain currents intersect). To reduce the large sense margin variation, the source degeneration scheme is used with longer channel length for PL transistors [2]. Source degeneration PMOS (PD) is added to the source of PL transistors to reduce current variation and increase effective resistance which result in SM improvement. Sense circuit is designed to reduce the impact of process variation on SM. This goal is achieved by increasing the width and length of PL transistors to reduce the $V_T$ mismatch between $PL_D$ and $PL_R$, and optimizing other design parameters (NC width, $V_{clamp}$ and $V_{Ref}$) to maximize both Sense-0 margin (SM0)and Sense-1 margin (SM1) (Fig. 2(b)). In addition, we have exploited clamp voltage as a knob to make a trade-off between SM0 and SM1 to minimize the sensing failures. In summary, the conventional sensing technique is fully optimized to ensure fair comparison with slope sensing technique.

### B. Basics of Slope Sensing Technique

Fig. 1(a) shows the proposed sensing circuit with features to inject the ramp current and sample the ramp voltage. The slope of

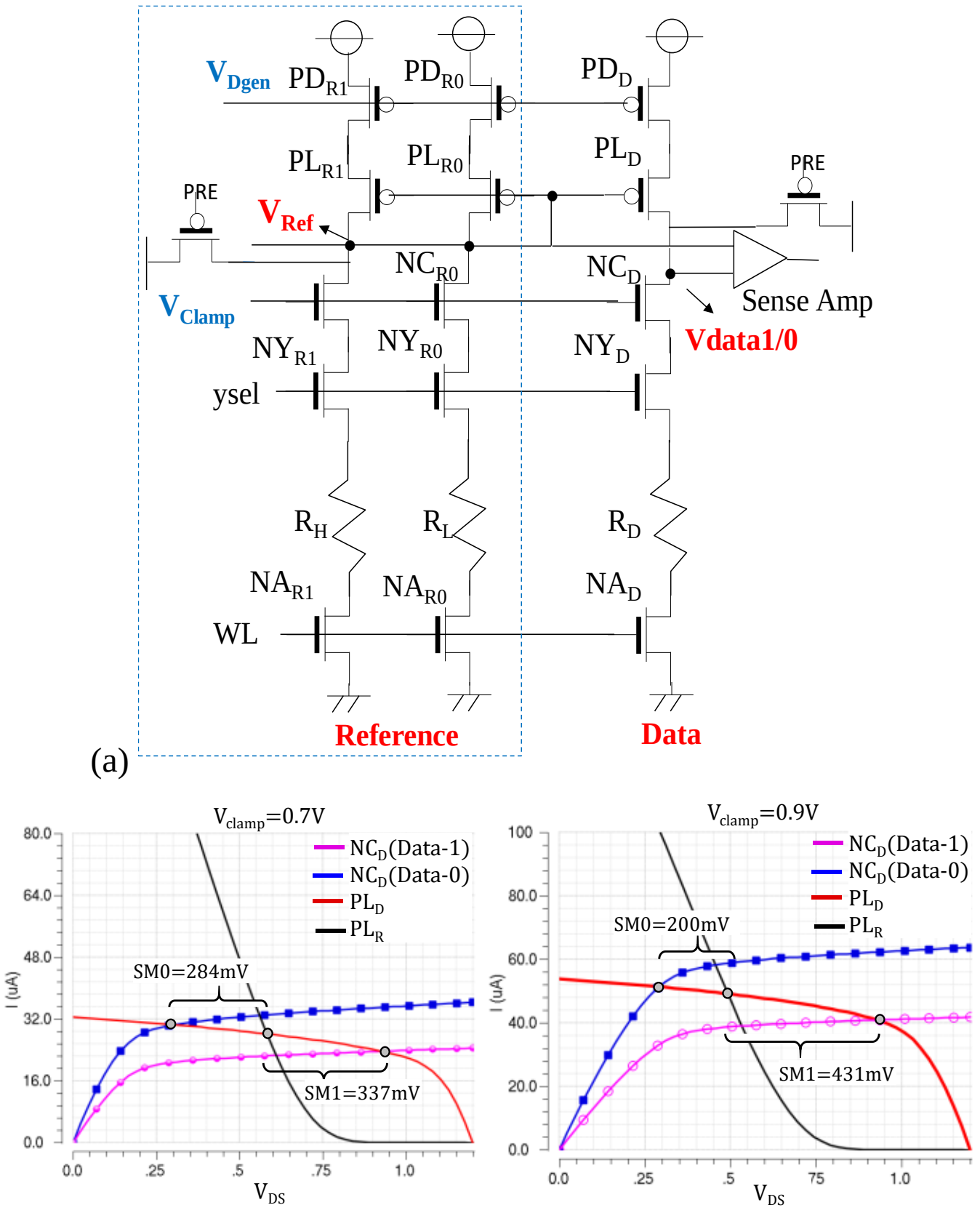

(b)

Fig. 2 (a) Conventional sensing circuit; and, (b) the impact of clamp voltage on sense margin for $V_{Clamp}$=0.7V and $V_{Clamp}$=0.9V.

ramp voltage will change from positive to negative if the MTJ switches from high to low resistance. The voltage slope remains positive if the MTJ resistance is initially low. The slope detection is performed by sampling the ramp voltage with two sample-and-hold circuits using clocks $\phi_1$ and $\phi_{1d}$ (delayed $\phi_1$). The sampled voltages are stored in $C_1$ and $C_{1d}$ respectively. Finally, $V_{C1}$ and $V_{C1d}$ are compared at the edge of sense amplifier enable (SE). Simplified timing diagram is shown in Fig. 1(b).

As shown in Fig. 3(a), white triangles are voltages sampled by $\phi_1$ and black triangle are voltages sampled by $\phi_{1d}$. The sense amplifier is enabled after the black triangles. As a result, two consecutive black and white triangle sampled voltages are compared. It is evident that the SM is positive in positive slope region and negative in the negative slope region. Sense margin depends on sampling frequency, slope of ramp current and MTJ switching time. Note that, MTJ switching time refers to the switching time of MTJ under a ramp current slope. The voltage across bitcell changes during switching from high to low resistance state. This voltage difference varies due to switching time variation resulting in sense margin variation. We have implemented design-for-test features to test the sensing failure by sweeping these parameters. By increasing the ramp current slope, the voltage difference between two consecutive samples will increase resulting in higher SM. However, the buffer output voltage ($V_{BUFO}$) may be clamped at $V_{DD}$ by further increasing the ramp current slope which results in SM loss.

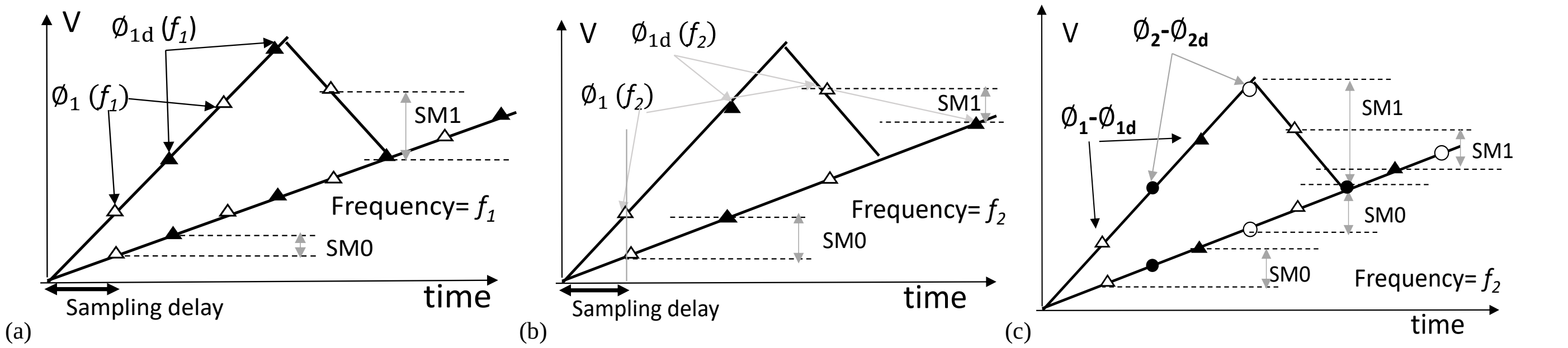

Fig. 3 Sampling voltage across MTJ: (a) sampling with frequency $f_1$ *and* $Ø_1$-$Ø_{1d}$ clock phases which provides poor SM0 and large SM1; (b) sampling with frequency $f_2$ ($f_2 = f1/2$) and $Ø_1$-$Ø_{1d}$ clock phases which provides large SM0 but poor SM1; and, (c) double sampling with frequency $f_2$, $Ø_1$-$Ø_{1d}$ and $Ø_2$-$Ø_{2d}$ clock phases which results in large SM0 and SM1 while ensure capturing negative slope.

### C. *Slope Sensing with Double Sampling*

Fig. 3 shows impact of two sampling frequencies on sense margin. Note that lower sampling frequency results in more SM0 since the voltage difference of two consecutive samples is higher as shown Fig. 3(b). However, decreasing the sampling frequency might cause an error in negative slope detection due to poor SM1. Sampling at higher frequency ensures negative slope detection. However, SM0 loss due to higher sampling frequency results in increased failures (Fig. 3(a)). As shown in Fig. 3(b), sampling with frequency $f_2$ (where $f_2 = f_1/2$) and $\phi_1$-$\phi_{1d}$ clock phases provides poor SM1 after MTJ flipping while sampling with frequency $f_1$, $\phi_1$-$\phi_{1d}$, provides larger SM1. In order to obtain the desired number of samples at lower sampling frequency to ensure negative slope detection (higher SM1) as well as higher SM0, double sampling technique is proposed.

Double sampling can be implemented by lowering sampling frequency and using two sets of sample-and-hold (S/H) with $\phi_1$-$\phi_{1d}$ and $\phi_2$-$\phi_{2d}$ clock phases (where $\phi_2$ and $\phi_{2d}$ are delayed version of $\phi_1$ and $\phi_{1d}$ respectively) to sample the voltage across the bitcell. Hence two groups of sample-and-hold circuits (SC) are used. From Fig. 3(c), sense amplifier is activated after $\phi_{2d}$ (black circle) and $\phi_{1d}$ (black triangles). Therefore, SM is the difference between black and previous white circle voltages which is sensed by the 1st sense amplifier or black and previous white triangle voltages which is sensed by 2nd sense amplifier. In the proposed double sampling method, if one of sense amplifiers detects negative slope (i.e., SM1) the output is '1' otherwise it is '0'. Therefore, the SM1 is the maximum absolute value of SM1 which is provided by two sets of S/H circuits. From Fig. 3(c), it can be noted that sampling with frequency $f_2$ and $\phi_1$-$\phi_{1d}$ provides poor SM1 while sampling with frequency $f_2$ and $\phi_2$-$\phi_{2d}$ provides large SM1. Therefore, double sampling with both $\phi_1$-$\phi_{1d}$ and $\phi_2$-$\phi_{2d}$ clock phases provides large SM1 and SM0. Sampling is performed during sense time ($T_{Sense}$) which ensures that all bitcells in high resistance switch to low resistance state under process variation. Note that sampling accuracy and robustness can be improved by increasing the number of sample-and-holds and lowering the sampling frequency to achieve larger SM1 and SM0 at the cost of more sense amplifiers.

## III. Test Chip Implementation

In this section, we explain the subarray architecture with integrated slope sensing circuit and the test chip design.

### A. *Slope Sensing Circuit Design*

Fig. 4 depicts the implementation details of slope sensing with two SCs to enable double sampling. To mimic MTJ resistance in the test-chip we used poly resistance. The bitcell contains two high resistances ($R_H$) and two access transistors (Fig. 4). The bitcell is in high resistance state if WL1 is activated and is in low resistance state if both WL1 and WL2 are asserted (two $R_H$ are connected in parallel). In order to mimic the switching time variation, we have incorporated a knob to fire WL2 at different times. Our design matches the real MTJ parameters such as, MTJ resistance, switching time and TMR variability using experimentally calibrated simulation models [10] (details in Section IIIB) and serves as a solid proof-of-concept for the slope sensing scheme.

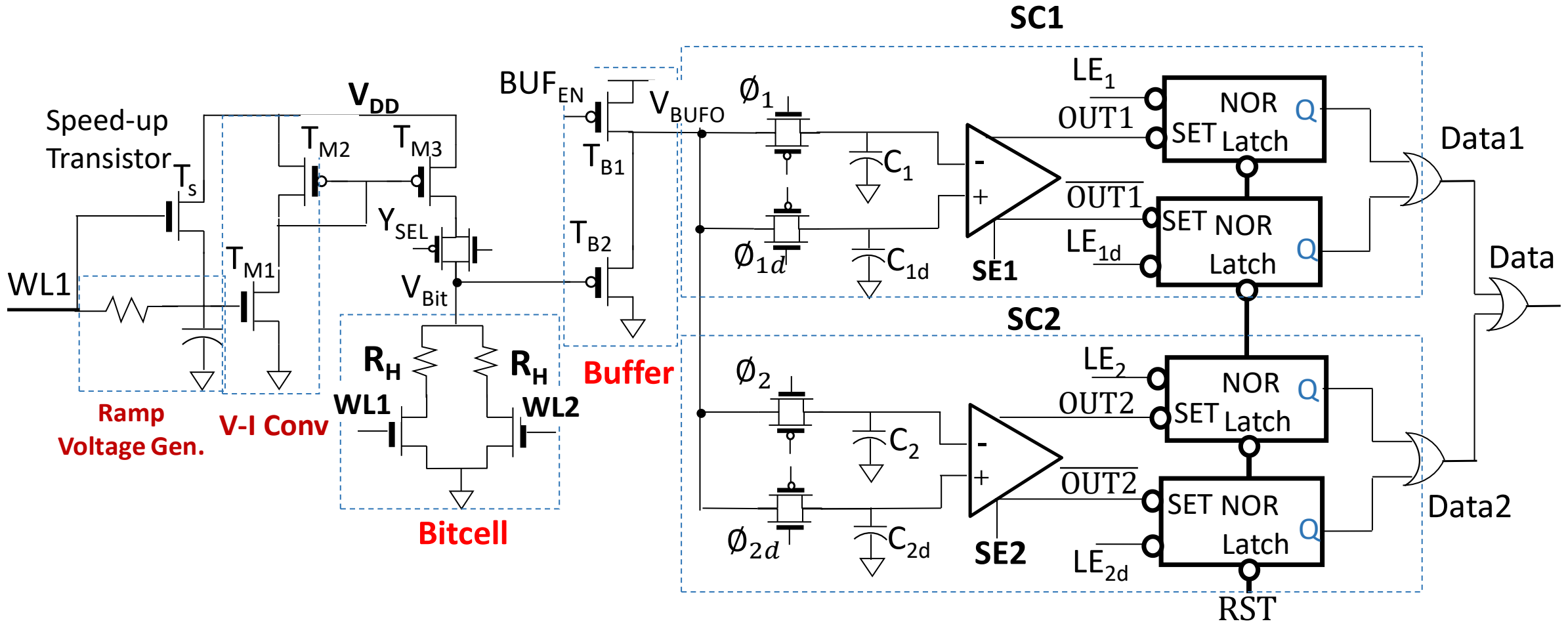

Fig. 4 Implementation details of slope detection sense circuit.

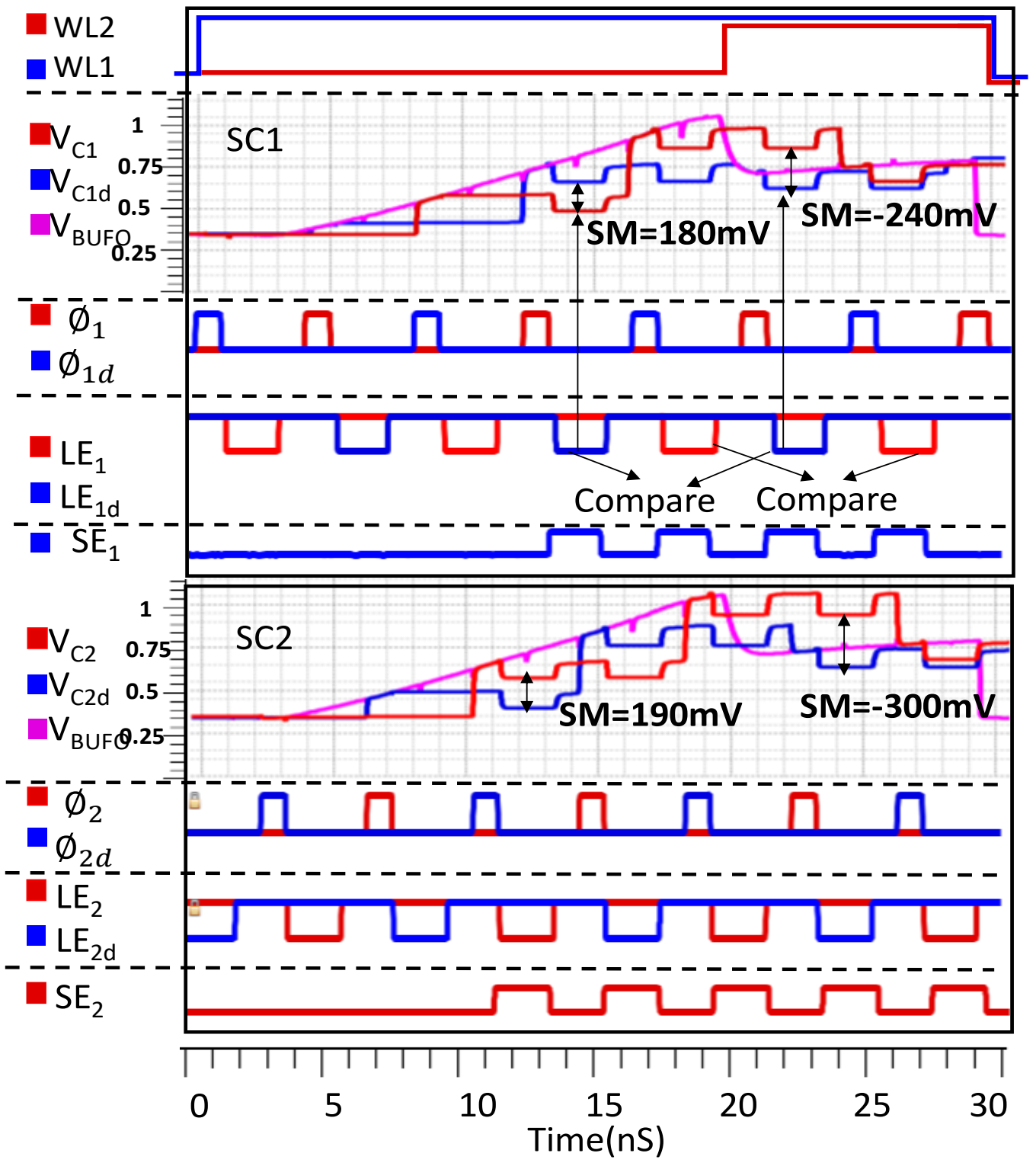

Fig. 5 Post layout simulation of slope sensing scheme along with timing diagram for sense circuit-1 (SC1) and SC2.

The ramp current is generated using RC low pass filter to generate a ramp voltage that is connected to gate of an NMOS transistor ($T_{M1}$) to generate a ramp current. Since the $T_{M1}$ is OFF for voltages less than threshold voltage, a speedup transistor ($T_S$) is used to charge the capacitor rapidly to threshold voltage of $T_{M1}$ in order to speed up the ramp current generation process. Next, the ramp current is injected in to bitcell using a PMOS current mirror which in turn generates a ramp voltage ($V_{BIT}$) at the input of PMOS source follower buffer ($T_{b2}$). In order to reduce the buffer offset voltage, width of $T_{b1}$ is larger than $T_{b2}$. Therefore, buffer offset is approximately the threshold voltage of $T_{b1}$. In order to reduce the offset voltage further, we use a transistor with low threshold voltage to provide enough headroom for output voltage swing. This reduces the buffer offset voltage to ~330mV.

Fig. 5 shows post layout simulation of slope sensing. To perform comparison between every two consecutive samples we exploited two NOR latches and two active-low latch enables ($LE_1$ and $LE_{1d}$) for each sense circuit. Comparison is performed at the SE edge. However, the comparison result is stored in the latches at the LE edge. Since OUT1 is connected to latch with $LE_1$, $V_{C1}>V_{C1d}$ (OUT is 0 when $V_{C1}>V_{C1d}$) indicates negative SM as a result a ‘1’ will be stored into latch (when OUT=0, ‘1’ will be latched). The latch with $LE_{1d}$ is connected to $\overline{OUT1}$, therefore, $V_{C1}<V_{C1d}$ indicates negative SM and output will be set to ‘1’. The outputs of two latches are ORed which indicates that the output is set to ‘1’ if one of the latches capture the negative slope. Bottom figure shows the timing diagram of SC2. The SC2 results in higher SM1 compared to SC1 (-240mV vs -300mV). If one of the sense circuits outputs ‘1’ the double sampling results in ‘1’ since outputs of two SCs are ORed. We designed a test feature to select between single and double sampling to study their impact on sensing errors.

Table-1: Parameters used for process-variation study

| Device | Parameter | Mean | Std. Dev. |
|---|---|---|---|
| MTJ | MgO Thickness | 1.2nm | 2.5% |
| MTJ (2.5k-10k) | Shape Area | 50*94 $nm^2$ | 15% |
| MTJ(5k-10k) | Shape Area | 30*94 $nm^2$ | 15% |

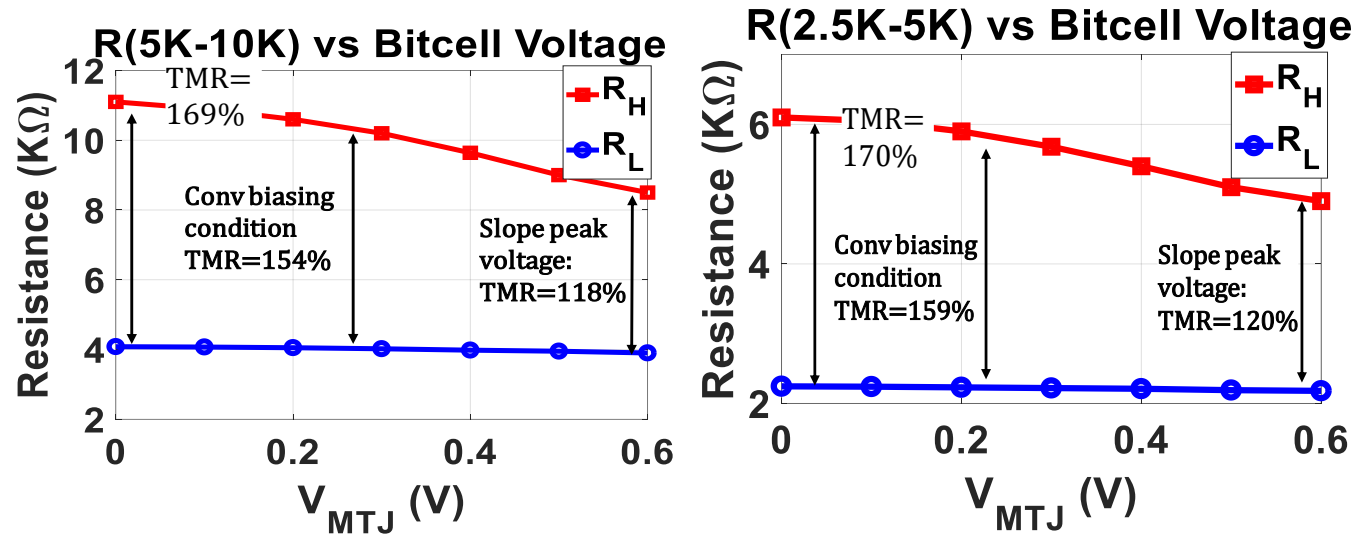

(a) (b)

Fig. 6 Dependency of TMR on applied voltage @120°C for (a) 5K-10K, and (b) 2.5K-5K MTJ resistances.

### *B. Impact of Process Variation on TMR and MTJ Resistance*

In order to mimic MTJ using poly resistance, Monte Carlo simulation is performed to characterize the behavior of MTJ under process variation. MTJ model [10] which is very well calibrated with experimental data is used to perform process variation analysis. For MTJ we have assumed tunnel oxide barrier thickness and surface area variations. The mean and standard deviation of these parameters are provided in Table-1. We consider MTJ area and oxide thickness variation reported in [6-7]. Different MTJ resistances is achieved by modifying the MTJ surface area (94nm*30nm for 5K-10K and 50nm*94nm for 2.5K-5K). MTJ is characterized in terms of switching time and resistance variation under slope and conventional sensing bias conditions by performing 4000 points Monte-Carlo simulations. In addition, we have investigated the dependency of applied voltage on TMR under basing conditions for both slope and conventional sensing. The TMR reduces with increased applied voltage across MTJ. Fig. 6(a-b) show the impact of applied voltage on TMR for both MTJ resistance flavors. TMR reduces 51% as applied voltage increases from 0V to 0.6V (peak voltage across MTJ in slope sensing) and 15% for

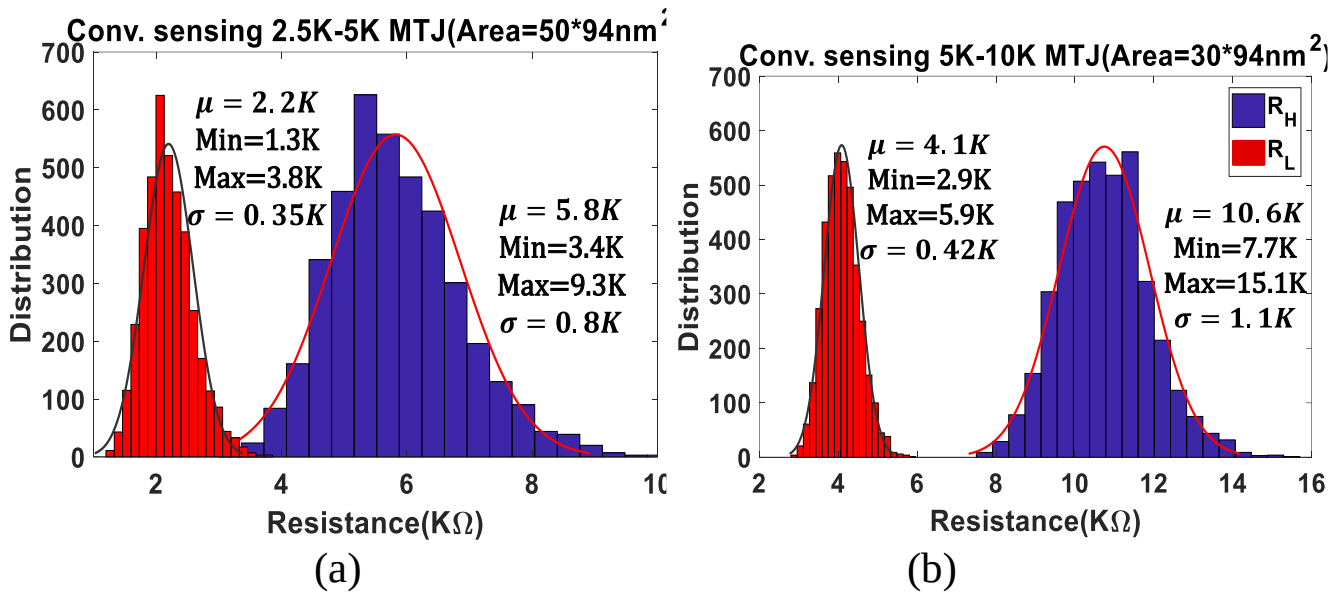

(a) (b)

Fig. 7 High and low resistance distribution under conventional sensing bias condition for 4000 points Monte Carlo simulation for, (a) 5K-10K, and (b) 2.5K-5K.

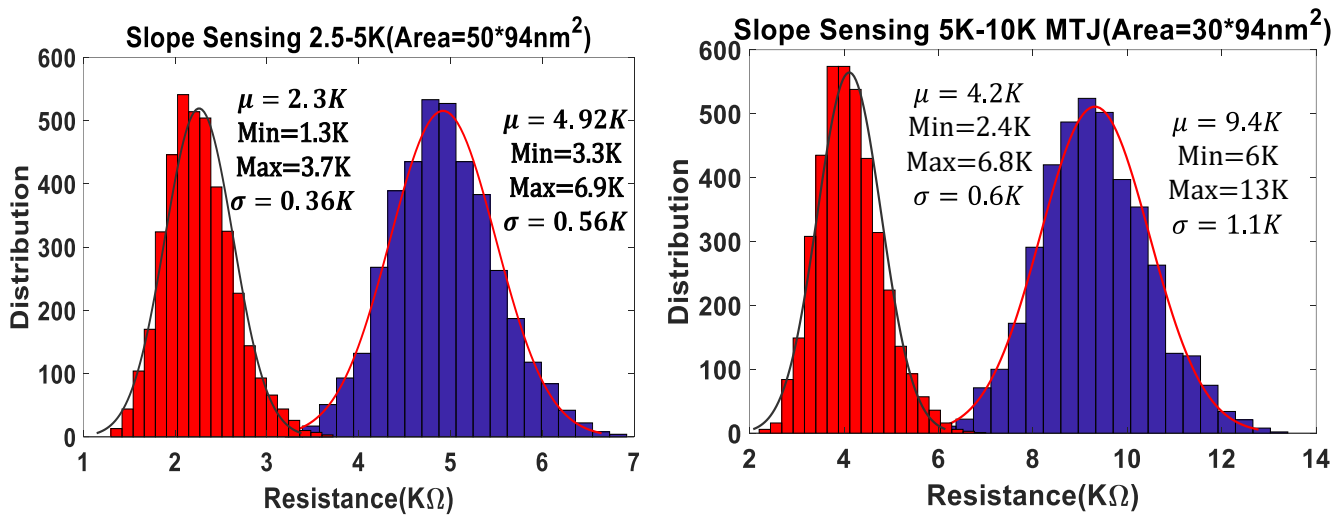

(a) (b)

Fig. 8 (a) High and low resistance distribution under slope sensing bias condition for 4000 points Monte Carlo simulation for, (a) 2.5K-5K, and (b) 5K-10K.

applied voltage of 0.25V (conventional sensing biasing condition).

Fig. 7 shows the $R_L$ and $R_H$ variation for two MTJ configurations (2.5K-5K and 5K-10K) under the conventional sensing bias condition for 4000 points Monte-Carlo simulation. It can be observed that one sigma of resistance variation is ~13% and ~11% for 2.5K-5K and 5K-10K MTJ configurations respectively. Conventional sensing achieves mean TMR of 150% and one sigma of 5%. Fig. 8 shows the MTJ resistance under slope sensing bias voltage (the peak voltage across MTJ is 0.6V). In case of 5k-10k, the mean of resistance in high and low resistance state is 9.4K and 4.9K respectively. One sigma of resistance variation in high and low resistance states are around 11% and 13%. It can be observed that one sigma of resistance variation is around 10% which matches the experimental results on MTJ variation reported in [8-9]. We have been very pessimistic in terms of TMR variation to make a fair comparison. As shown in Fig. 9, the mean of TMR is 128% and 120% and one sigma of the TMR variation is 8% and 10% for 5k-10k and 2.5k-5k MTJ flavors respectively (we considered a pessimistic guard band of $\sim u - 6\sigma$ for TMR). We incorporated test feature to modulate the TMR. For slope sensing, the TMR can be varied from 80% to 120% and for conventional sensing the TMR can be varied from 60% to 120%. The resistance can be varied by -10%, -20% and +10%, +20% to mimic the MTJ resistance variation as well as impact of TMR on sensing failures. In addition, the poly resistance varies due to process variation. The poly resistance variation using 1000 points MC simulation is shown in Fig. 10. One sigma of poly resistance variation is ~8%. Moreover, poly resistance shows 15% variation across process corners. The poly resistance variation itself accounts for mimicking MTJ variation. We believe that poly resistance fluctuation can model both TMR and process variation in MTJ.

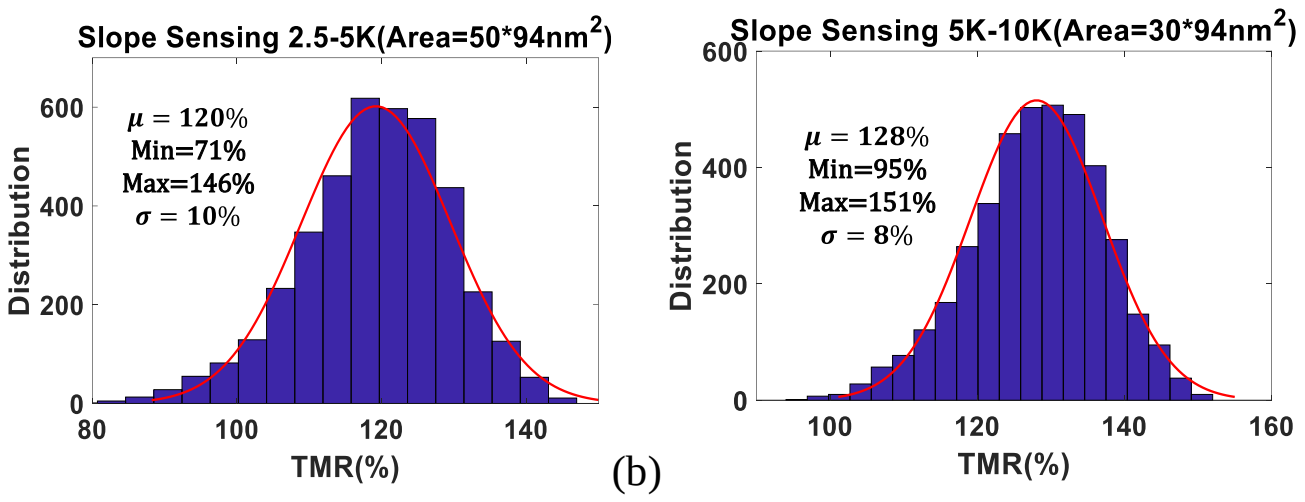

(a) (b)

Fig. 9 TMR distribution under slope sensing bias condition for 4000 points Monte Carlo simulation for, (a) 2.5K-5K, and (b) 5K-10K.

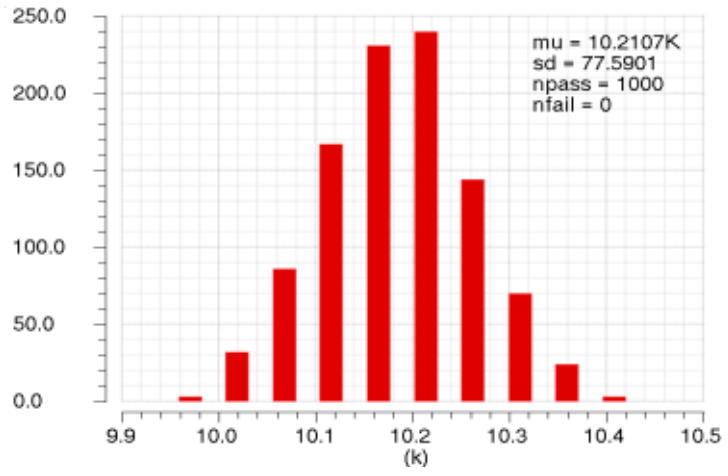

Fig. 10 Poly resistance intra-die process variation.

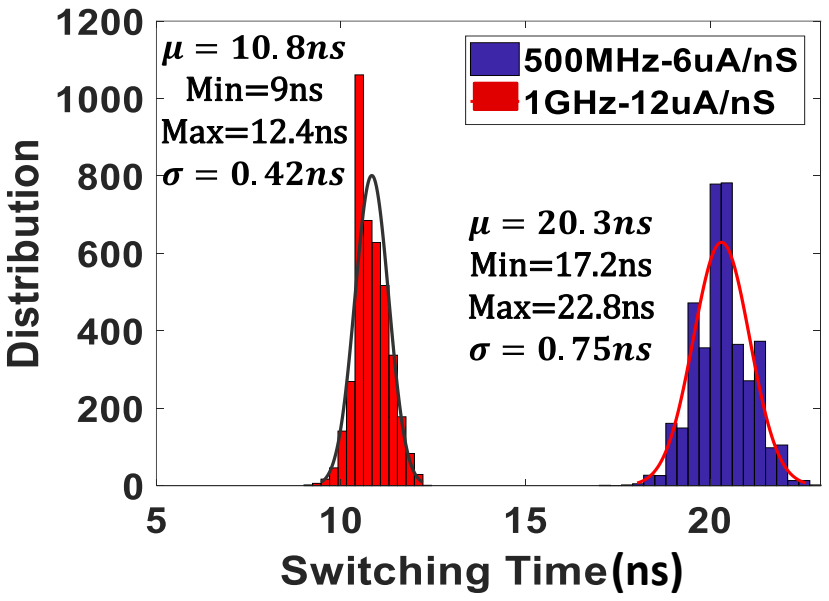

Fig. 11 MTJ switching time distribution for 6uA/ns and 12uA/ns ramp current slopes for 1000 Monte Carlo points for 2.5K-5K MTJ config.

The MTJ switching time variation is obtained by performing 4000 points Monte Carlo simulation for different ramp current slope. The ramp current is applied as long as WL1 is active. Since the WL1 pulse and sampling clocks are generated using system clock, faster clock frequency results in shorter WL1 period (sensing duration) and higher sampling speed. As mentioned earlier, the SM degrades with increased sampling frequency. This demands for higher ramp current slope to achieve large sense margin. Therefore, at lower system clock of 500MHz, ramp current with slope of 6uA/ns is applied while for higher clock frequency of 1GHz the slope of ramp current is 12uA/ns to achieve higher SM as shown in Fig. 11. Based on simulation model which captures both stochastic switching due to random thermal fluctuation and process variation, the switching time varies from 17.2ns to 22.8ns with $\mu = 20.3ns$ and $\sigma = 0.75ns$ for applied ramp current of 6uA/ns. Test feature is incorporated to vary switching time from 16ns to 24ns (8 to 12 cycles). Fig. 11 depicts the MTJ switching time for 6uA/ns and 12 uA/ns ramp current slopes. Note that we have been very conservative in terms of switching time variation, and modulated switching time by $\sim u \pm 6\sigma$. The MTJ switching time

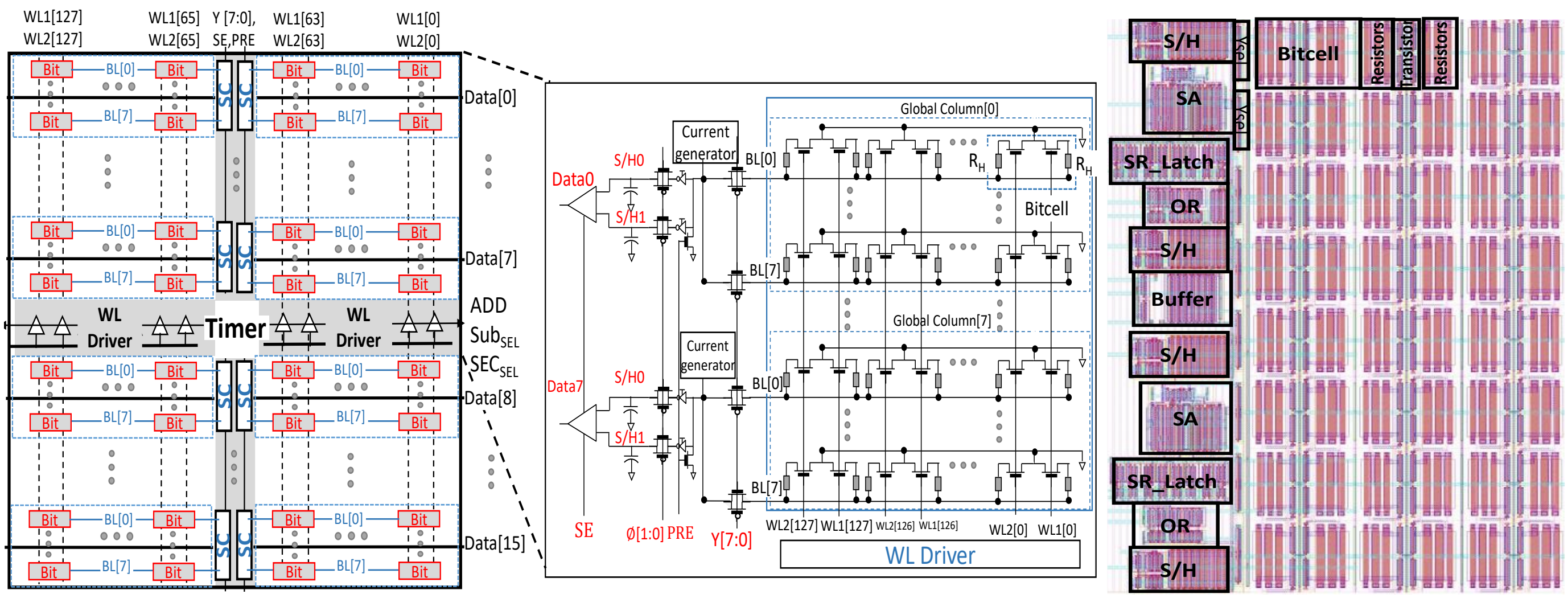

(a) (b)

Fig. 12 (a) Subarray architecture. The sector architecture is shown in inset; and, (b) the right sector layout. The column circuitry is also shown.

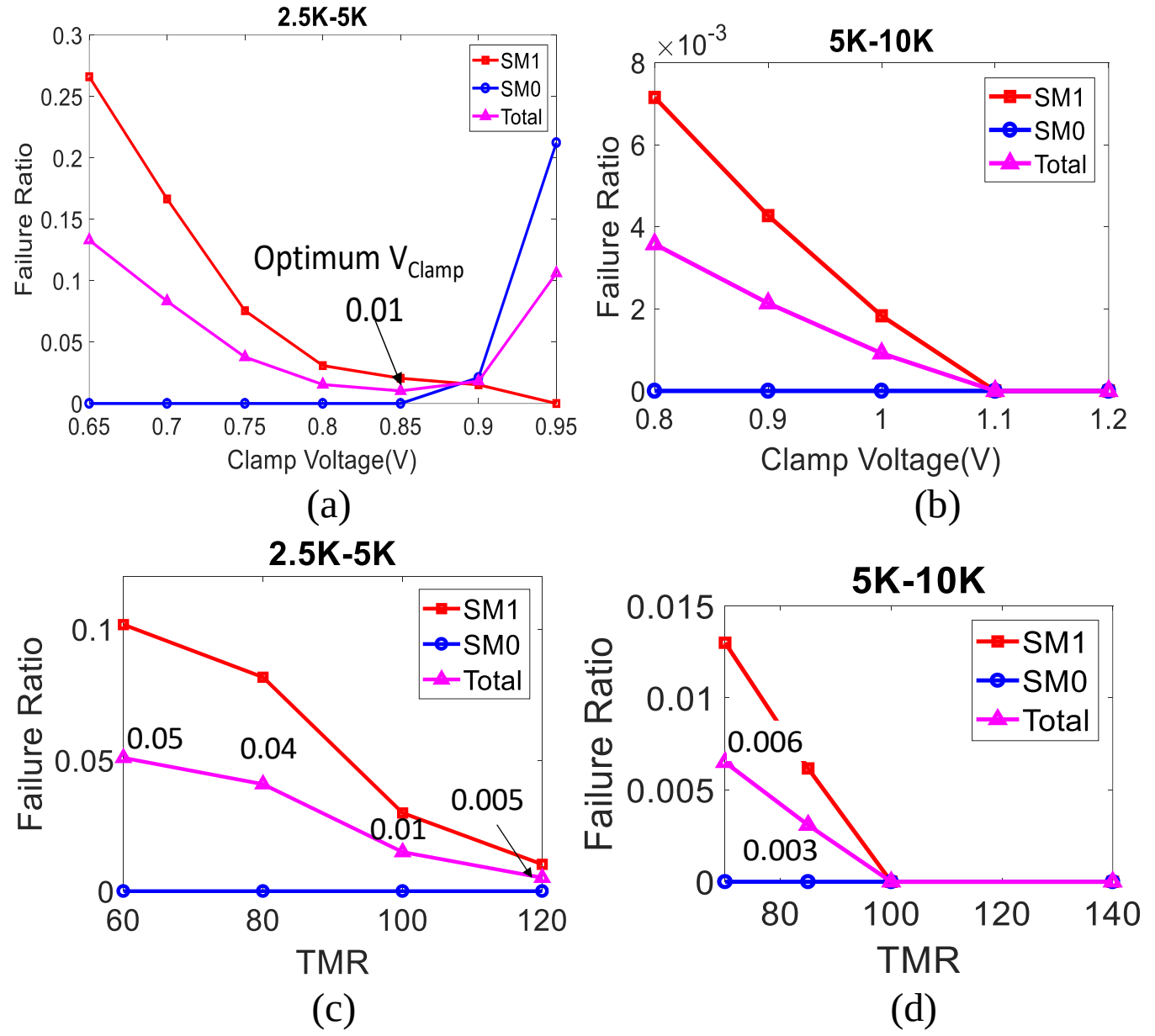

Fig. 13 Experimental results: (a)-(b) Conventional sensing failure ratio with respect to clamp voltage for 2.5K-5K and 5K-10K arrays for TMR of 100%; and, (c)-(d) failure ratio with respect to TMR for 2.5K-5K and 5K-10K arrays with optimum clamp voltage.

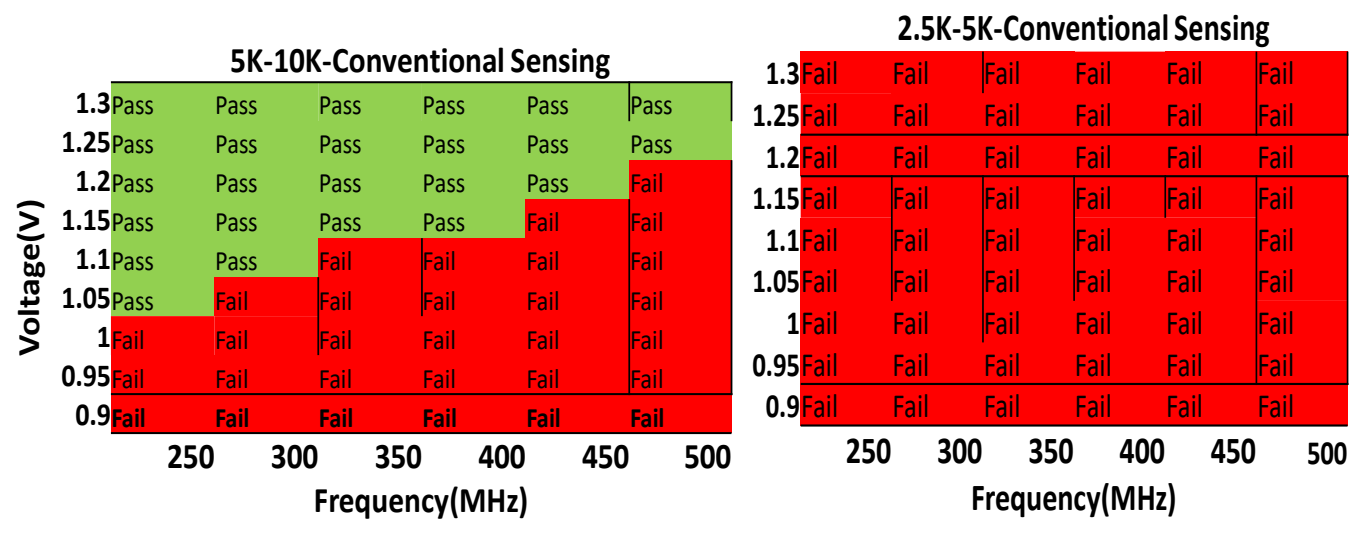

(a) (b)

Fig. 14 Experimental results: Conventional sensing shmoo plot with TMR of 100% and optimum clamp voltage for (a) 5K-10K array; and, (b) 2.5K-5K array.

is modulated by enabling the WL2 at various clock cycles to mimic MTJ switching time variation.

Note that, even though the test-chip does not contain the real MTJ, it can characterize the sensing failure under switching time variation. In this manner failures due to poor sense margin can be characterized and isolated from other failures such as MTJ defect (since the MTJ technology is not mature yet), BL-BL short and BL open. Although this test-chip cannot completely mimic the MTJ, it provides an opportunity to characterize sensing failure with respect to different flavors of MTJ resistance and TMR which is not possible to investigate using real MTJ.

### C. *Array Architecture*

Fig. 12 (a) shows the array architecture. To study the effect of resistance value on sensing we implemented array of 2.5K/5K and 5K/10K MTJ resistances. The resistances are tunable by +/- 20% to explore the effect of TMR variation on sensing errors. A partial layout of right sector containing six WLs and column circuitry is shown in Fig. 12(b). In this test-chip, the bitcell layout is bigger than that of array of real MTJ due to use of poly resistor. However, the size of sense circuit is matched with the size of access transistor (which is the size of real MTJ array) as shown in Fig. 12(b). The sense circuit layout can be easily matched with the size of sense circuit by modifying the column multiplexing and access transistors aspect ratio without affecting the conclusions/results drawn in this paper. To characterize slope sensing we swept clock frequency from100Mhz to 500MHz (sampling frequency is 1/4$^{th}$ of the clock frequency), ramp current slope (5 to 14 uA/ns) and MTJ switching time (8 to 12 cycles). For conventional sensing, we swept clamp voltage ($V_{clamp}$).

## IV. Test Results

In this section, first we explain the conventional sensing experimental result and the impact of TMR and clamp voltage on the sensing failures. Next, we describe the experiential results that presents the impact of ramp current slope, sampling frequency, switching time on sensing failure. Moreover, we depict the shmoo plot as well as impact of process variation on sensing failures. Finally, we compare the conventional sensing failures against slope sensing.

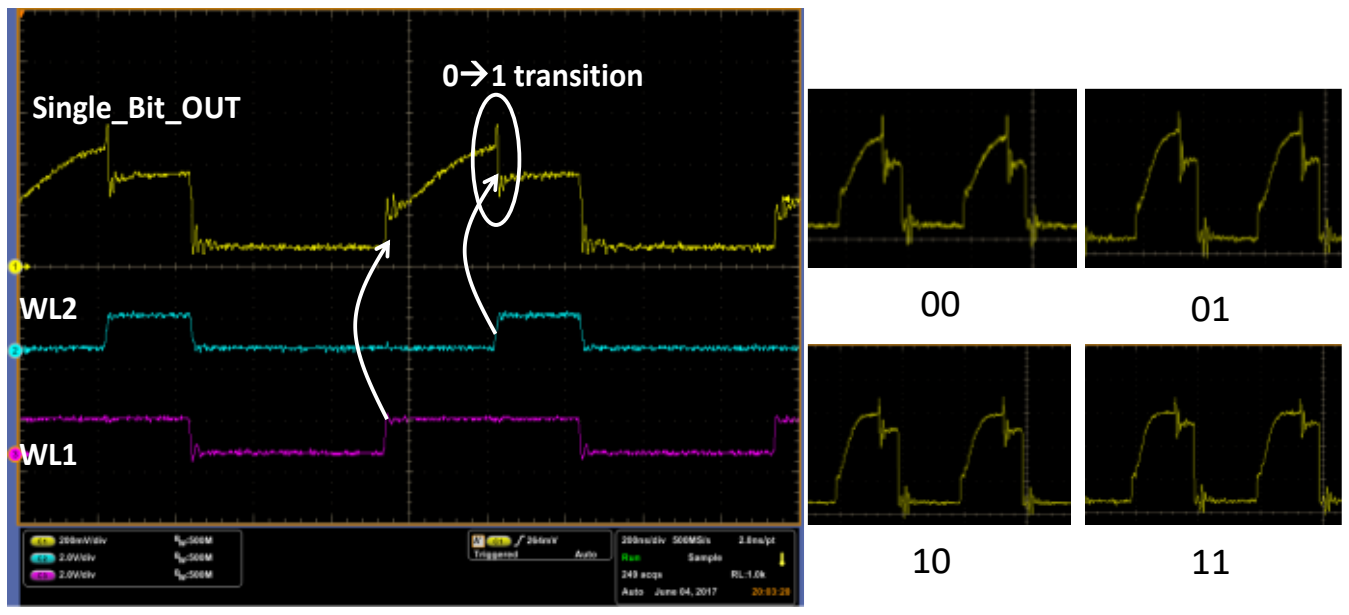

(a) (b)

Fig. 15 Oscilloscope capture of voltage across single-bitcell. Sensing starts by activating WL1 and bitcell switches to low resistance state at the edge of WL2; and, (b) the slope of voltage across bitcell for various current slope settings. Setting 00 indicates the lowest and 11 indicates the highest current slope.

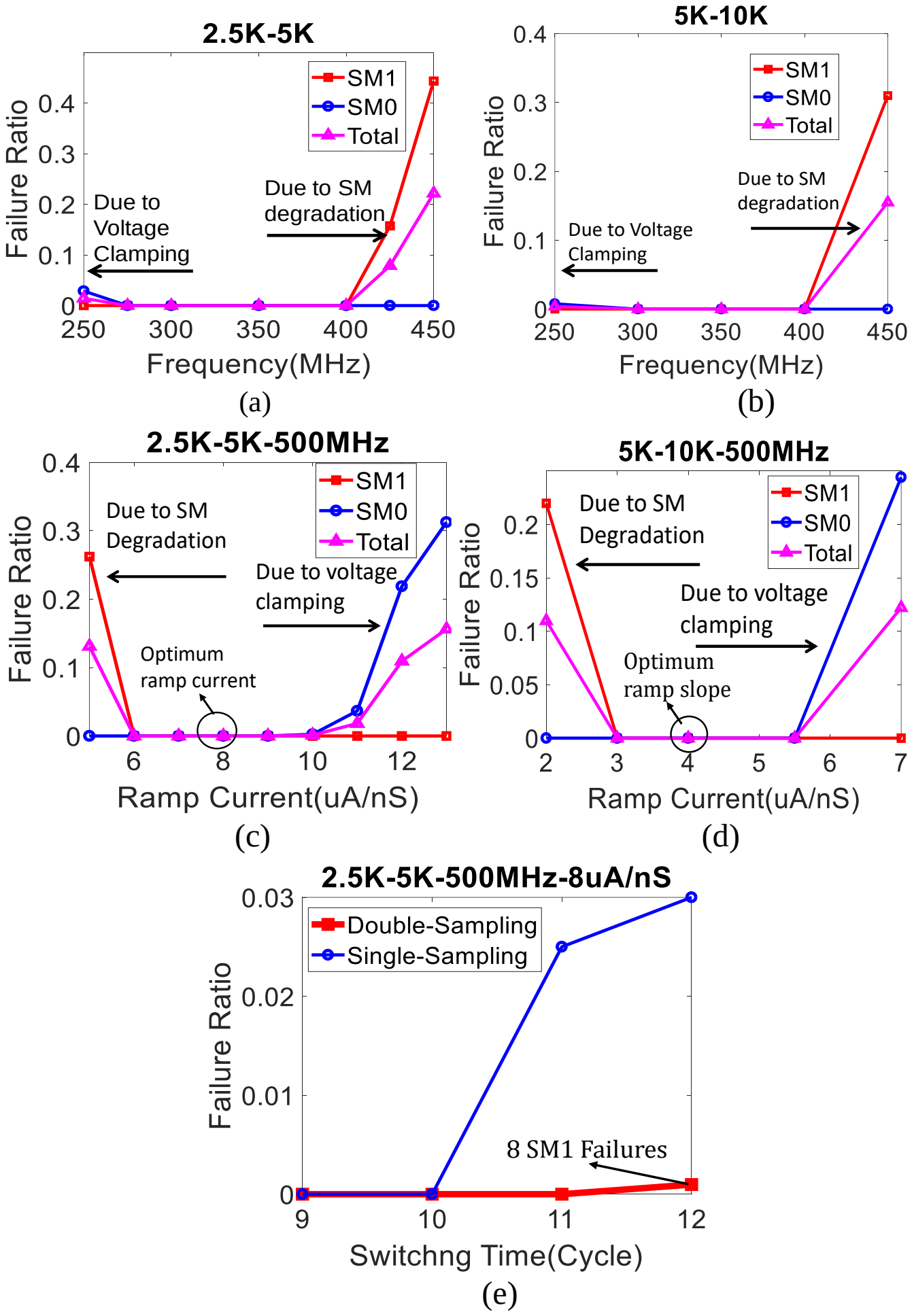

Fig. 16 Experimental results: (a)-(b) Slope sensing failure ratio with clock frequency for 2.5K-5K and 5K-10K arrays; (c)-(d) failure ratio with ramp current slope for 2.5K-5K and 5K-10K arrays; and, (f) failure ratio with switching time for double and single sampling method.

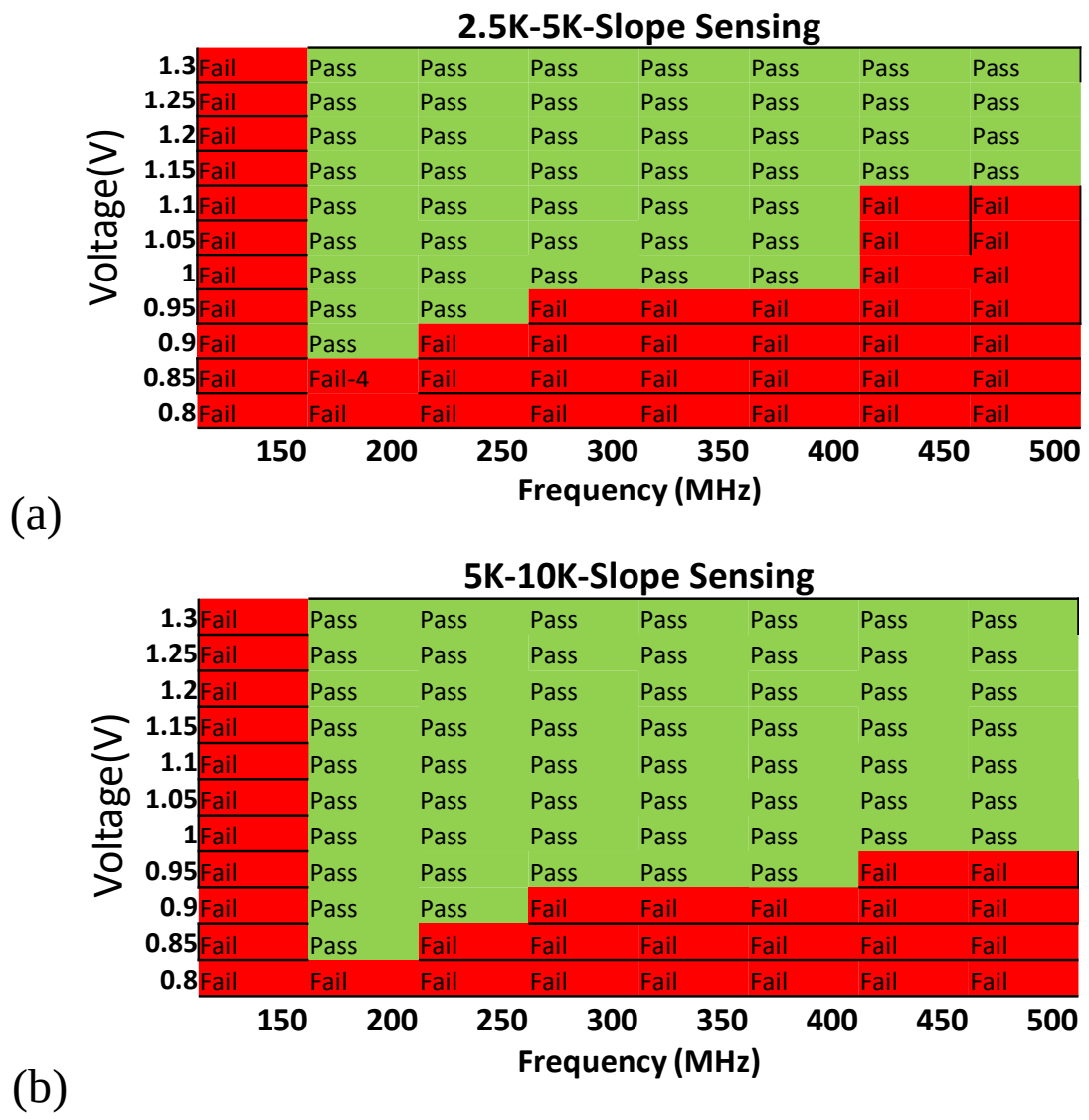

Fig. 17 Experimental results: Slope sensing shmoo plot with TMR of 100% and optimized ramp current slope and double sampling for, (a) 2.5K-10K array; and, (b) 5K-10K array. The # of failing chips out of 10 tested chips for failing voltage and frequency is shown.

### A. *Conventional Sensing Test Results*

Fig. 13(a)-(b) shows the conventional sensing failures vs $V_{clamp}$ for 2.5K-5K and 5K-10K at TMR=100%. By increasing $V_{clamp}$, the SM0 failures increases while SM1 failures decreases. This plot trend matches the simulation results discussed in Section IIA. For 2.5K-5k array the optimum $V_{clamp}$ = 0.85V which result in minimum failure ratio (=0.01). The 5K-10K array results in zero failure ratio at $V_{clamp}$ =1.1V. This is due the higher difference between low/high resistance and reference resistance for 5K-10K than 2.5K-5K array. Fig. 13(c)-(d) shows failure ratio vs TMR. The 5K-10K array performs better than 2.5K-5K array. Fig. 14(a)-(b) shows the shmoo plots of 5K-10K and 2.5K-5K arrays with 100% TMR. Note that the 2.5K-5K array fails for all frequency and voltages.

### B. *Slope Sensing Test Results*

To demonstrate slope sensing, we designed a proof of concept single bitcell that works at low frequency (the WL period is 500ns) to capture the high-to-low switching waveforms. Sensing starts by activating WL1 and bitcell switches to low resistance state at the edge of WL2 result in negative slope (Fig. 15(a)). Fig. 15(b) shows the slope of voltage across bitcell for various current slope settings. Setting '00' ('11') provides lowest (highest) current slope. Thus, the negative slope can be captured to determine memory state. Single bitcell contains large source follower buffer at the output to drive the large capacitance of output analog pads.

Fig. 16(a)-(b) shows the array-level slope sensing failures vs clock frequency for 2.5K-5K and 5K-10K arrays. Note that sampling frequency is one fourth of clock frequency for each S/H circuit. Lower than 250MHz clock result in failures due to voltage clamping. Due to longer WL at slower clock for constant ramp current slope, the peak voltage across bitcell will increase and can get clamped at $V_{DD}$ leading to SM loss. More than 400 MHz clock results in sensing failures due to SM1 loss because of sampling at higher frequency. Fig. 16(c)-(d) shows the failure ratio vs ramp current slope. In the case of 2.5K-5K array, the failures increase for ramp current slope lower than 6uA/ns due to SM loss. Ramp current slope greater than

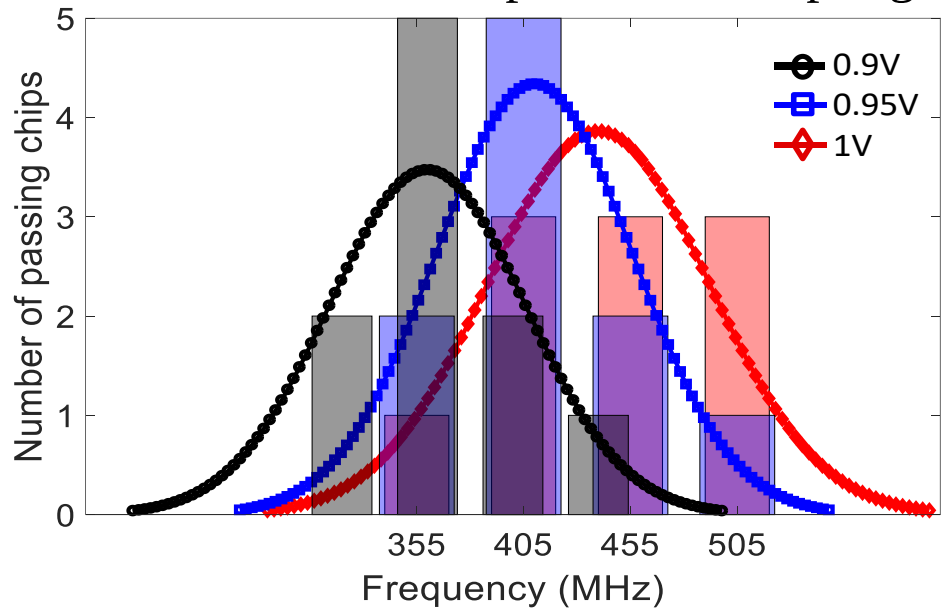

Fig. 18 Experimental results: Passing frequency distribution for 10 tested chips for 2.5K-5K array.

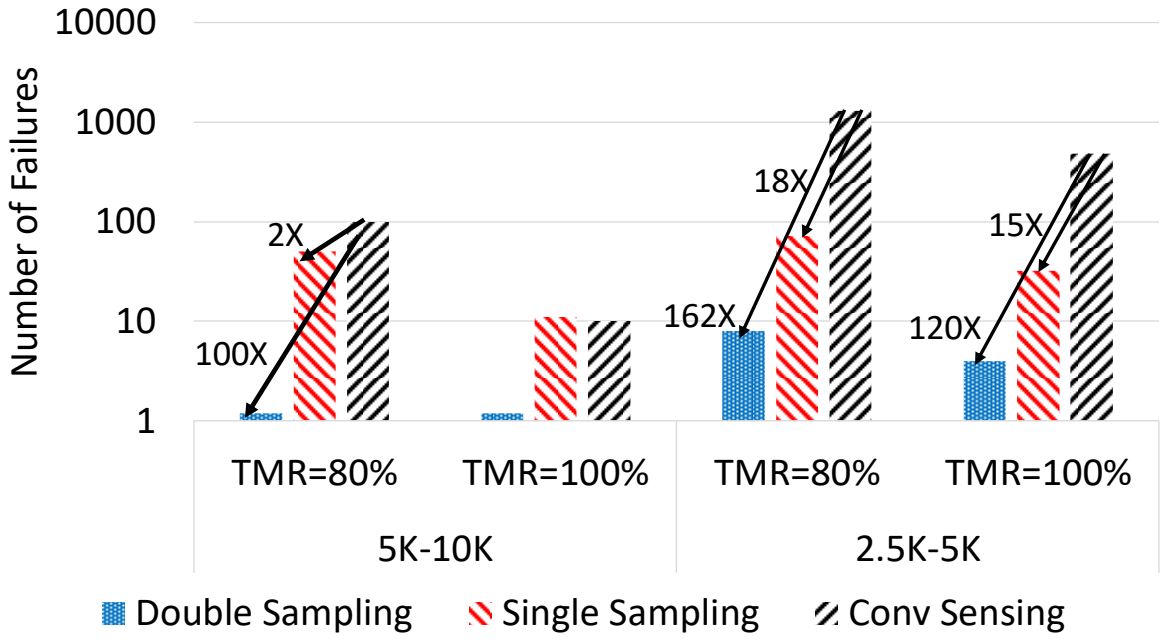

Fig. 19 Experimental results: Comparison of # of failures for conventional and slope sensing.

10uA/ns result in sense failures due to voltage clamping. Since MTJ switching time changes significantly due to process variation we have swept it by changing the WL2 assertion time. Fig. 16(e) shows the failure ratio with respect to switching time for double and single sampling method for 2.5K-5K array at 500MHz. It can be observed that double sampling method reduces the SM1 failures significantly under MTJ switching time variation. Fig. 17(a)-(b) shows the shmoo plot for 2.5K-5K and 5K-10K arrays @ TMR = 100%. Note that slope sensing results in zero error for wide voltage and frequency range. To study the effect of process variation, we have tested 10 chips and plotted the passing frequency for 1V, 0.95V and 0.9V (Fig. 18). The passing frequency increases for higher voltage.

Fig. 19 shows the comparison of slope and conventional sensing. Slope sensing results in 100X failure reduction for TMR=80% in 5K-10K array and 120X (162X) failure reduction for TMR=100% (80%) in 2.5K-5K array. Fig. 20 shows chip microphotograph and design features.

We have compared the proposed sensing with state-of-art sensing techniques (Table-2). Even though slope sensing read latency is higher and consume more power, it provides higher read yield in presence of process variation. 2T2MTJ are proposed in [15] and [17] to implement reference-less sensing. Although these techniques can achieve low read latency of 3.3ns and 4ns respectively, employing additional MTJ and access transistor reduces the bitcell density significantly. Furthermore,

Table-2: Comparison with other sensing schemes.

| Sensing Technique | Techn-ology | Supply Voltage | Capacity | Power (uW) | Read-Out Time(nS) | Average SM | Failure rate | Reference less |
|---|---|---|---|---|---|---|---|---|
| Slope Sensing (this work) | 65nm | 1V-1.2V | 96Kb | 190 | 26@500 MHz | 200mV (2.5K-5K) | 0% | ✓ |
| Conventional Sensing (this work) [1-2] | 65nm | 1V-1.2V | 96Kb | 90 | 13@500 MHz | 180mV | 1% (2.5K-5K) 0% (5k-10K) | - |
| Self-Reference [5] | 240nm | 2V | 16Kb | - | 130 | ~40mV | 0% | ✓ |
| Non-Dest. Self-Reference [4] | 130nm | 1.2V-1.5V | 16Kb | ~100 | 15 | ~20mV | 2% | ✓ |
| SPSC (Simulation)[11] | 45nm | 1V | - | 33.5 | 3 | 600mV | Read Yield=5.7σ | - |
| VFAB(Simulation) [12] | 65nm | 1.2V | - | 16.2 | 5.2 | 800mV | Read Yield=9.8σ | - |
| ISSCC '09 2T1MTJ [14] | 90nm | 1.5 | 32Mb | - | 12 | - | - | - |
| ISSCC'15 2T2MTJ [15] | 65nm | 1.2 | 1Mb | 71.2 | 3.3 | - | - | ✓ |
| ISSCC'10 [16] | 65nm | 1.2 | 64Mb | 83.16 | 20 | - | - | - |
| VLSI '13 2T2MTJ [17] | 65 | 1.2 | 1Mb | 75 | 4 | >50mV | - | ✓ |
| JSSC '18 [18] | 28 | 1.2/1.8 | 1Mb | 230 | 2.8 | - | $10^{-5}$ | - |

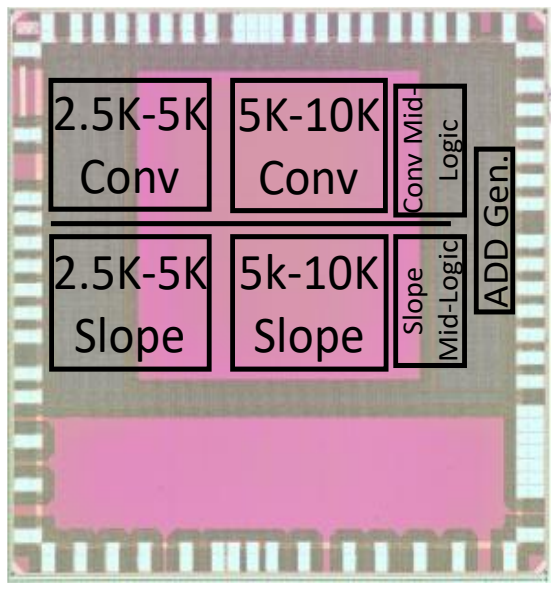

| Technology | 65nm ST Microelectronics |
|---|---|
| Capacity Conventional sensing | 96Kb (6* 16Kb) |
| Capacity slope sensing | 96Kb (6* 16Kb) |
| Subarray (16Kb) | 256*128 |
| Die size | 2mm*2mm |
| Array Size | 0.5mm*0.5mm |
| Supply Voltage | 1.2V |
| Slope Sensing read time | 26nS@500MHz |
| Conv sensing read time | 13nS@500MHz |

Fig. 20 Chip microphotograph and features.

two MTJs must be written during write operation which increase the write power by a factor of two.

However, the proposed slope sensing offers high density compared to reference-less 2T-2MTJ. Furthermore, it achieves lower area (~18% savings in this test-chip) compared to the conventional sensing [2] due to elimination of reference bitcells at the expense of higher latency and power overhead. We have not reported area since real MTJ has not been used. In conventional sensing each global column contains 8 data bitlines and 2 reference bitlines.

## V. Discussion

### A. *Sense Amplifier Offset Voltage Analysis*

The Sense Amplifier (SA) offset voltage depends on sense time and sense amplifier size since increasing transistor size decreases the transistor threshold voltage variation. We design the sense amplifier in such a way to reduce the offset while meet the area and delay requirements. We considered sense time of 0.5ns. In order to obtain sense amplifier offset, the Vref is provided to one of sense amplifier input and the Vdata is swept. For each voltage sweep 1000 points Monte-Carlo simulation is performed and tabulated sense amplifier failure distribution is shown in Fig. 21. This distribution can be modeled by a Gaussian distribution with $\mu = 8$mV and $\sigma = 16$mV.

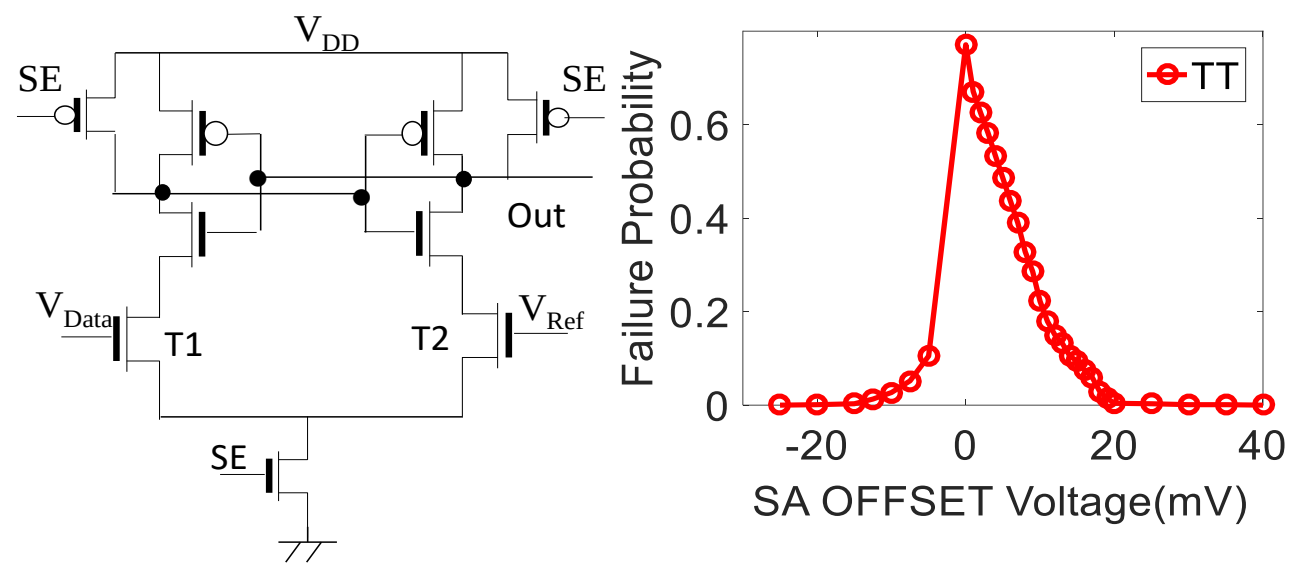

(a) (b)

Fig. 21(a) Sense amplifier circuit; and, (b) SA offset voltage distribution for 1000 points Monte-Carlo simulations.

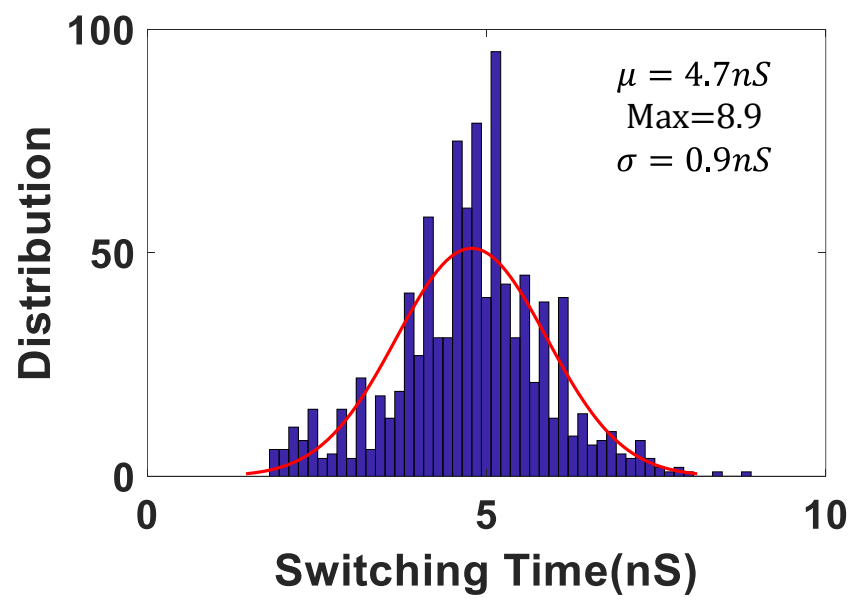

Fig. 22 Write latency distribution for 5000 Monte Carlo points.

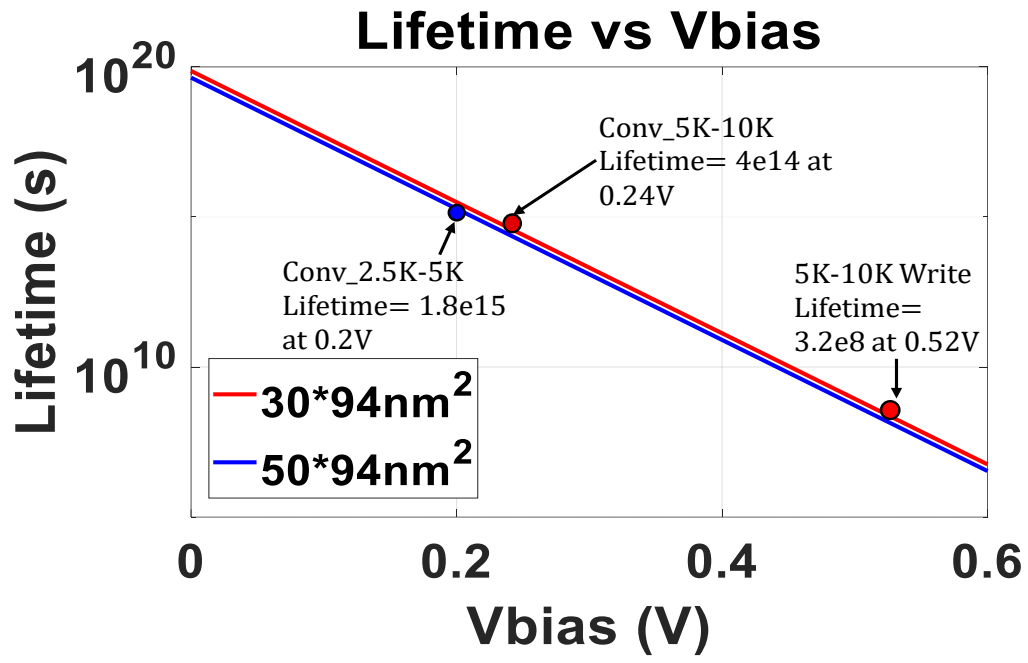

Fig. 23 Estimated lifetime using E-model for varying bias voltage for different values of junction area for 2.5K-5K and 5K-10K MTJs.

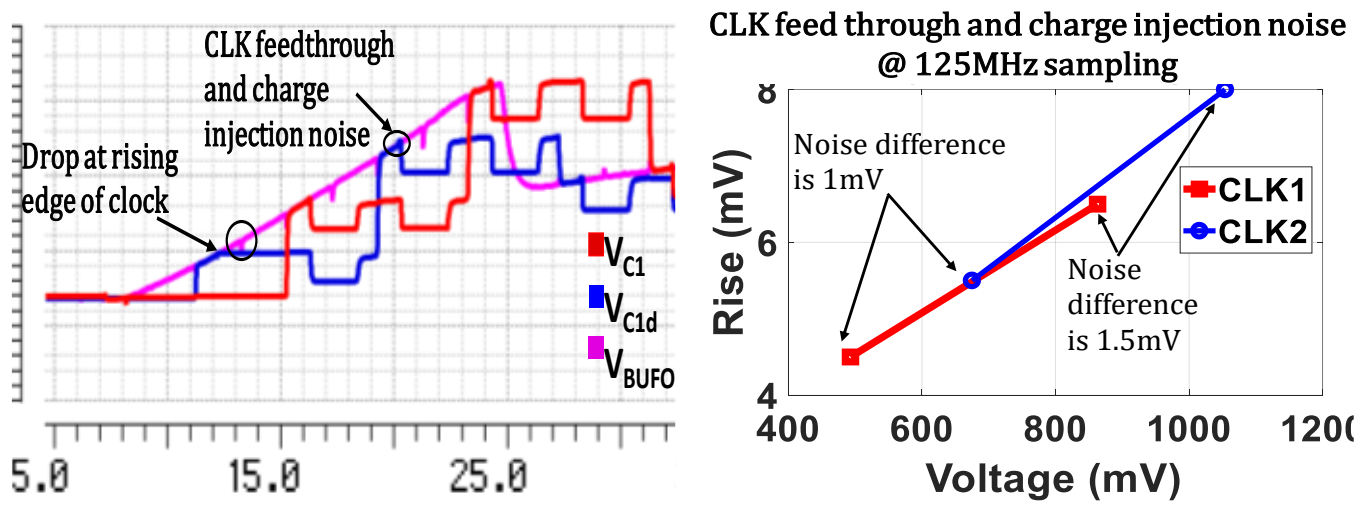

(a) (b)

Fig. 24 (a) Noise induced on $V_{BUFO}$ and output sampled voltage; and, (b) noise due to charge injection and CLK feedthrough at falling edge of clock.

## B. *Write Back Operation*

The read operation is performed in 13 cycles (26ns @500MHz). The MTJ switching during read operation is slow due to small applied ramp current. However, the write operation can be completed faster by applying higher current. We have performed monte-carlo simulation in order to capture switching time variation. The write latency is asymmetric in nature. Therefore, we have considered the worst-case polarity (high→low transition) for latency analysis. As shown in Fig. 22 the mean write latency is 4.7ns for write current of 100uA. To ensure the all bitcell are written properly we consider a margin of $\mu+6\sigma$=10ns for write pulse duration. Write-back can start immediately after negative slope detection and only the bitcells storing logical '1' must be written back. The write back operation can be overlapped with address decoding and precharge stage or during the idle cycles of memory where the data is sent to the edge of bank.

## C. *MTJ endurance analysis*

Lifetime of MTJs is usually measured with respect to the Time Dependent Dielectric Breakdown (TDDB) mechanism. Read and write operations create voltage drop across the MTJ. The thin oxide barrier experiences high electric field which degrades the reliability of the device. The applied voltage induces an electric field which causes an overall reduction of activation energy $E_A$, leading to an increased breakdown probability. A detailed analysis is presented in [19] to establish that the observed intrinsic dielectric breakdown of the oxide barrier follows the E-model. According to the E-model the breakdown probability is defined as:

$$p(t) = \frac{dF(t)/dt}{1-F(t)} \quad (1)$$

Where F(t) denotes the fraction of devices that break intrinsically after a time 't' and p(t) is defined as:

$$p(t) = A\exp\left(\frac{V(t)}{B}\right) \quad (2)$$

In the case when dV /dt is a constant, an intrinsic failure F(t) can be given as:

$$F(t) = 1 - \exp\left[-p(t)B\left(\frac{dv}{dt}\right)^{-1} + AB\left(\frac{dv}{dt}\right)\right] \quad (3)$$

For a time independent breakdown probability density p(t), the mean lifetime is expressed as:

$$\tau_{1/2} = \frac{\ln(2)}{p(t)} \quad (4)$$

where $\tau_{1/2}$ is the time for when 50% of the devices experience breakdown. By curve fitting the experimental data from Ref. [20], we derived the corresponding values of constants for A and B to be $9.43*10^{-21}$ and 0.019 respectively assuming a constant rate of degradation due to both reads and writes. Fig. 23 shows MTJ lifetime for $T_{OX}$=1.2nm and cross-sectional area of $30*94nm^2$ (5K-10K), and $50*94nm^2$(2.5K-5K). The voltage across MTJ is higher in high resistance state which results in worst-case lifetime. The voltage across MTJ in conventional 5K-10K array is 240mV for worst-case 10K resistance and the voltage across MTJ in 2.5K-5K array is 200mV for worst-case 5K resistance. As shown in Fig. 23, 5K-10K MTJ life time is $4*10^{14}$ while 2.5K-5K lifetime is $1.8*10^{15}$. In conventional sensing technique WL is high for 12n. Hence, $8.3*10^7$ read operation can be performed in one second which results in endurance of $24*10^{21}$ for 5K-10K array and $11*10^{22}$ for 2.5K-5K array using conventional method.

In case of slope sensing, the slope of voltage with respect to time (dV/dt) is 0.6V/20ns (the pick of voltage across MTJ is 0.6V).

The lifetime of MTJ can be calculated using Eq. 3 with dV/dt=0.6V/20ns which results in lifetime of $1.8*10^{10}$ and $1.9*10^{10}$ for 2.5K-5K array and 5K-10K array respectively. In slope sensing a ramp voltage is applied for 26ns across MTJ. Hence, in one second $3.8*10^{7}$ read operation can be performed under stress imposed by slope sensing biasing condition. This results in endurance of $6.84*10^{17}$ and $7.2*10^{17}$ for 2.5K-5K and 5K-10K array respectively. Therefore, the slope sensing degrades the endurance ~$10^{4}$ times.

### D. Sampling Noise Analysis

We have performed noise analysis. Fig. 24(a) shows the drop in $V_{BUFO}$ at rising edge of the sampling clock. The voltage which is stored in the sampling capacitors (~3fF) results in a drop in $V_{BUFO}$ at rising edge of sampling clock. However, the buffer is strong enough to charge the capacitor in 40ps while the sampling clock pulse width is 0.5ns. Therefore, this noise will not appear at the output sampled voltage. The noise due to charge injection and clock feedthrough appears as a rise in sampled voltage at the falling edge of clock since this technique employs differential sampling. This noise is negligible as shown in Fig.24(b). The sampled voltage using CLK1 ($V_{C1}$) is compared against that of CLK2 ($V_{C2}$). Therefore, the effective noise is the difference between noise induced by first and second sample which is very small and in order of 1mV as shown in Fig. 24(b).

## VI. Conclusions

We have designed a test-chip to demonstrate reference-less slope sensing technique to eliminate bit-bit variations. We characterized the slope sensing failures with respect to ramp current slope, sampling frequency and various resistance values. A 96kb fabricated test-chip in 65nm technology shows that slope sensing reduces failure rate by 120X in 2.5K-5K array@TMR=100% and 162X in 2.5K-5K@TMR=80% array compared to conventional voltage sensing.

## Acknowledgements

We thank SRC 2727.001 and DARPA D15AP00089 for funding. We also thank Dr. Jaydeep Kulkarni for initial discussions and Nitin Rathi for help in layout.